\documentclass[journal]{IEEEtran}
\usepackage{cite}
\usepackage{amsmath,amssymb,amsfonts}
\usepackage{graphicx}
\usepackage{algorithm}
\usepackage{algpseudocode}
\usepackage{array}
\usepackage{multirow}
\usepackage{url}
\usepackage{xcolor}
\usepackage{microtype}
\newtheorem{lemma}{Lemma}
\newtheorem{proposition}{Proposition}
\newtheorem{remark}{Remark}
\newcommand{\ind}{\mathbf{1}}

\begin{document}

	\title{\huge{Joint Trajectory and Power Optimization for Movable-Antenna-Aided Secure Transmission}}
\author{Jiacheng Liao, Fengchao Zhu, Guojie Hu, Ying Zhang and Tong-Xing Zheng, \textit{Senior Member, IEEE}
		\vspace{-5pt}	
	\thanks{J. Liao, F. Zhu (corresponding author),  G. Hu and Y. Zhang are with the College of Communication Engineering, Rocket Force University of Engineering, Xi'an, 710025, China (email: jiachengliao2024@163.com, fengchao\_zhu@126.com,  lgdxhgj@sina.com, ying\_zhang2025@126.com).

		T.-X. Zheng is with the School of Information and Communications Engineering, Xi'an Jiaotong University, Xi'an, 710049, China (email: zhengtx@mail.xjtu.edu.cn).
}}
	\maketitle
	\setlength{\abovedisplayskip}{4.5pt plus 1.5pt minus 1.5pt}
	\setlength{\belowdisplayskip}{4.5pt plus 1.5pt minus 1.5pt}
	\setlength{\abovedisplayshortskip}{2.5pt plus 1pt minus 1pt}
	\setlength{\belowdisplayshortskip}{2.5pt plus 1pt minus 1pt}
\begin{abstract}
This paper investigates joint trajectory and power optimization for movable antenna (MA)-aided secure transmission, where the MA can transmit during its movement.
Specifically, the average secrecy rate is maximized subject to movement and power constraints, which couple the MA positions and transmit powers over time, respectively.
To tackle this nonconvex problem, we first decouple the transmit powers via Lagrangian duality, where the optimal power for a given trajectory is derived in semi-closed form and shown to depend only on the current MA position.
Accordingly, we decouple the MA positions by equivalently reformulating the problem on a discretized linear rail in terms of the farthest grid point and the number of time slots spent at each grid point.
Then, from strong duality of the convex relaxation for each farthest point, we construct an asymptotically optimal solution in which the MA dwells at no more than two positions.
Finally, a low-complexity dual-based branch-and-bound algorithm is developed to select the farthest grid point and certify the optimality gap.
Simulation results show that the proposed design outperforms the benchmark schemes.

\end{abstract}


	\section{Introduction}

\IEEEPARstart{P}{hysical}-layer security (PLS) has emerged as a promising paradigm to safeguard wireless communications by exploiting the intrinsic randomness of wireless channels~\cite{Wyner1975,Gopala2008}.
Recently, movable antennas (MAs)~\cite{Zhu2024} have been exploited to enhance PLS~\cite{Hu2024,Tang2025,Ding2025}, where the antenna positions are optimized within a confined region to enlarge the legitimate channel's advantage over the eavesdropping channel.
However, these works neglect the MA movement toward the optimized positions. In practice, an MA moves at a finite speed~\cite{Wei2026}, and thus its movement incurs a non-negligible delay that reduces the time available for data transmission~\cite{QLi2026}.

To exploit the MA movement period, a joint communication and trajectory design was proposed in~\cite{Li2026}, where the trajectory of a single MA is optimized via dynamic programming (DP) to maximize the average communication rate.
Although this design can be extended to maximize the average secrecy rate~\cite[Remark~1]{Li2026}, its transmit power remains fixed.
However, the secrecy rate is zero wherever the eavesdropping channel is dominant, regardless of the transmit power.
Under an average power constraint, the power saved at such positions can thus be reallocated to more favorable ones~\cite{Gopala2008}.
To the best of our knowledge, joint trajectory and power optimization remains unexplored for MA-aided PLS.

In this paper, we maximize the average secrecy rate by jointly optimizing the trajectory and transmit power of an MA that can transmit while moving.
 By dualizing the average power constraint, we show that the optimal power for a given trajectory depends only on the current MA position and vanishes in both the Eve-dominant and budget-driven zones.
Accordingly, the problem on a discretized linear rail is equivalently reformulated in terms of the farthest grid point and the number of time slots spent at each grid point.
Based on the convex relaxation for each farthest grid point, we construct an asymptotically optimal solution, where the MA dwells at no more than two positions and otherwise moves at the maximum speed.
To select the farthest grid point, a low-complexity dual-based branch-and-bound (BnB) algorithm is developed, in which dual upper bounds prune unpromising candidates and certify the optimality gap.
Simulation results demonstrate that the proposed design outperforms the benchmark schemes.


	\section{System Model and Problem Formulation}\label{sec:model}

	\begin{figure}[!t]
		\centering
		\includegraphics[width=0.8\columnwidth]{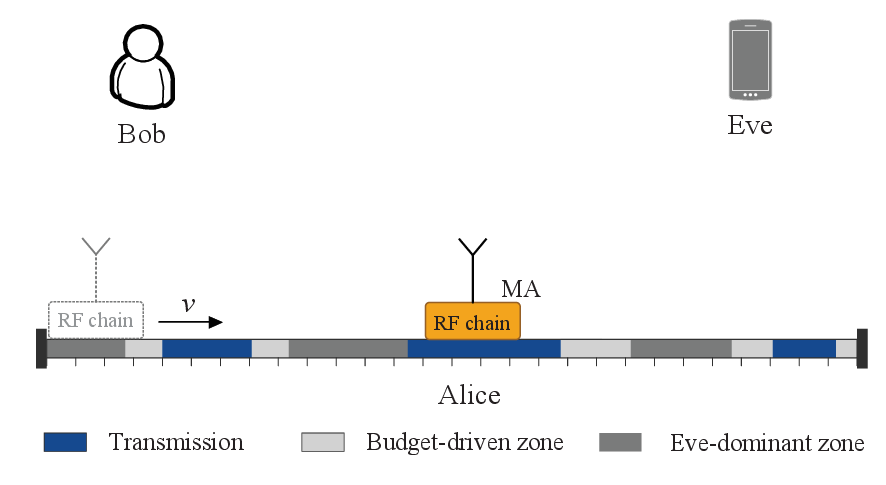}
		\caption{MA-aided secure transmission system, where the rail colors indicate whether the MA transmits or stays silent (see Section~\ref{sec:power}).}
		\label{fig:sys}
	\end{figure}

	As illustrated in Fig.~\ref{fig:sys}, a transmitter (Alice) sends confidential messages to a legitimate receiver (Bob) in the presence of an eavesdropper (Eve). Alice is equipped with a single MA that moves along a linear rail $\mathcal{C}\triangleq[0,D]$, whereas Bob and Eve are each equipped with a single fixed-position antenna.
   The propagation environment is assumed to be quasi-static within a block of duration $T$, which is divided into $K$ time slots of length $\tau\triangleq T/K$, indexed by $k\in\mathcal{K}\triangleq\{1,\dots,K\}$.
     With $v_{\max}$ and $\lambda$ denoting the maximum speed of the MA and the carrier wavelength, respectively, $K$ is assumed sufficiently large to satisfy $v_{\max}\tau\ll\lambda$, so that the channel remains approximately constant within each time slot~\cite{Li2026}.
      Let $x[k]\in\mathcal{C}$ denote the MA position in time slot $k$, where the MA starts from a given initial position $x[0]=x_0$.

Under the field-response channel model~\cite{Zhu2024}, the channel from Alice to node $u\in\{\mathrm{b},\mathrm{e}\}$ (with \(\mathrm{b}\) and \(\mathrm{e}\) referring to Bob and Eve, respectively) at MA position $x\in\mathcal{C}$ is given by
\begin{equation}\label{eq:chan}
	h_u(x)=\sum_{l=1}^{L_u}a_{u,l}e^{\,j\kappa_{u,l}x},\qquad \kappa_{u,l}\triangleq\frac{2\pi}{\lambda}\cos\theta_{u,l},
\end{equation}
 where $L_u$ is the number of paths, and $\theta_{u,l}\in[0,\pi]$ and $a_{u,l}$ are the angle of departure (AoD) and complex gain of the $l$-th path, respectively. In time slot $k$, Alice transmits a confidential symbol ${c}[k]$ with $\mathbb{E}\{|{c}[k]|^2\}=1$ at power $p[k]$, and the signal received at node $u$ is
\begin{equation}\label{eq:rx}
	y_u[k]=h_u(x[k])\sqrt{p[k]}\,{c}[k]+n_u[k],
\end{equation}
where $n_u[k]\sim\mathcal{CN}(0,\sigma_u^2)$ is the additive white Gaussian noise. The received signal-to-noise ratios at Bob and Eve are thus $\alpha(x[k])p[k]$ and $\beta(x[k])p[k]$, respectively, where the channel-to-noise ratios are defined as
\begin{equation}\label{eq:margin}
	\alpha(x)\triangleq\frac{|h_{\mathrm{b}}(x)|^2}{\sigma_{\mathrm{b}}^2},\qquad
	\beta(x)\triangleq\frac{|h_{\mathrm{e}}(x)|^2}{\sigma_{\mathrm{e}}^2}.
\end{equation}
Following~\cite{Li2026,Hu2024}, we assume Alice has perfect knowledge of $h_u(x)$ for all $x\in\mathcal{C}$.\footnote{This assumption is reasonable when Eve is an untrusted user in the network, whose path parameters \(\{\theta_{\mathrm{e},l},a_{\mathrm{e},l}\}\) can be estimated by Alice via MA channel estimation~\cite{Ma2023CE} and channel reciprocity.}
Accordingly, the instantaneous secrecy rate at position $x$ with transmit power $p$ is given by
\begin{equation}\label{eq:rate}
	r(x,p)\triangleq\left[\log_2\frac{1+\alpha(x)p}{1+\beta(x)p}\right]^{+},
\end{equation}
where $[z]^{+}\triangleq\max\{z,0\}$. Note that for $p>0$, $r(x,p)>0$ if and only if $\Delta(x)\triangleq\alpha(x)-\beta(x)>0$.
Under peak and average power budgets $P_{\max}$ and $\bar P$, respectively, the average secrecy rate maximization problem is formulated as
	\begin{subequations}\label{eq:P1}
		\begin{align}
			(\mathrm{P1}):\ &\max_{\{x[k],p[k]\}}\ \frac{1}{K}\sum_{k\in\mathcal{K}}r\big(x[k],p[k]\big)\label{eq:P1obj}\\
			&\ \ \mathrm{s.t.}\ \ |x[k]-x[k-1]|\le v_{\max}\tau,\ \forall k,\label{eq:P1speed}\\
			&\hphantom{\ \ \mathrm{s.t.}\ \ }x[0]=x_0,\ \ x[k]\in\mathcal{C},\ \forall k,\label{eq:P1rail}\\
			&\hphantom{\ \ \mathrm{s.t.}\ \ }0\le p[k]\le P_{\max},\ \forall k,\label{eq:P1peak}\\
			&\hphantom{\ \ \mathrm{s.t.}\ \ }\frac{1}{K}\sum_{k\in\mathcal{K}}p[k]\le\bar P,\label{eq:P1avg}
		\end{align}
	\end{subequations}
where~\eqref{eq:P1speed} and~\eqref{eq:P1rail} are the movement constraints, and~\eqref{eq:P1peak} and~\eqref{eq:P1avg} are the peak and average power constraints, respectively.
Problem (P1) is challenging to solve due to the nonconcavity of $r(x,p)$ with respect to (w.r.t.) $x$ and the time coupling imposed by~\eqref{eq:P1speed} and~\eqref{eq:P1avg}.
	Nevertheless, we next derive the optimal power structure, based on which (P1) is decoupled over time through an equivalent reformulation.

	\section{Problem Reformulation}\label{sec:struct}
	\subsection{Power Allocation for a Given Trajectory}\label{sec:power}
To reveal the optimal power structure, we first fix the trajectory $\{x[k]\}$, so that (P1) reduces to a power allocation problem over $\{p[k]\}$.
For brevity, we omit the argument $x$ of $\alpha(x)$, $\beta(x)$, and $\Delta(x)$ in this subsection.
For $\Delta>0$, the derivative of $r(x,p)$ w.r.t. $p$ is
\begin{equation}\label{eq:fprime}
	\frac{\partial r(x,p)}{\partial p}=\frac{\Delta}{(1+\alpha p)(1+\beta p)\ln 2},
\end{equation}
which is positive and strictly decreasing w.r.t. $p$, whereas $r(x,p)=0$ for $\Delta\le0$.
Hence, $r(x,p)$ is nondecreasing and concave w.r.t. $p$, and this power allocation problem is convex, for which the Karush--Kuhn--Tucker (KKT) conditions are necessary and sufficient~\cite{Boyd2004}.

Let $\nu\ge0$ be the dual variable for~\eqref{eq:P1avg}.
Since~\eqref{eq:P1avg} is the only constraint that couples the time slots, the Lagrangian $\frac{1}{K}\sum_{k\in\mathcal{K}}\big[r(x[k],p[k])-\nu p[k]\big]+\nu\bar P$ is separable over time slots.
 Hence, it can be maximized in each time slot by solving the following problem at $x=x[k]$, whose optimal value is termed the net secrecy gain at $x$:
\begin{equation}\label{eq:wdef}
	w(x;\nu)\triangleq\max_{0\le p\le P_{\max}}\{r(x,p)-\nu p\},
\end{equation}
i.e., the secrecy rate minus the power cost priced by $\nu$.

Since~\eqref{eq:fprime} equals $\Delta/\ln2$ at $p=0$, the objective function of~\eqref{eq:wdef} is nonincreasing w.r.t. $p$ if $\Delta\le\tilde\nu\triangleq\nu\ln2$, which yields $p=0$ and $w(x;\nu)=0$.
Otherwise, the stationarity condition $(1+\alpha p)(1+\beta p)=\Delta/\tilde\nu$ admits a unique positive root
\begin{equation}\label{eq:pcirc}
	p^{\circ}(x;\nu)=\frac{2(\Delta/\tilde\nu-1)}{\alpha+\beta+\sqrt{(\alpha-\beta)^2+4\alpha\beta\Delta/\tilde\nu}},
\end{equation}
with $p^{\circ}(x;0)\triangleq\infty$. Projecting it onto $[0,P_{\max}]$ yields the optimal solution to~\eqref{eq:wdef} as
\begin{equation}\label{eq:pstar}
p^{\star}(x;\nu)=\min\{P_{\max},p^{\circ}(x;\nu)\}\,\ind_{\{\Delta>\tilde\nu\}},
\end{equation}
where $\ind_{\{\cdot\}}$ denotes the indicator function.
Since $p^{\circ}(x;\nu)$ strictly decreases from $\infty$ to $0$ as $\tilde\nu$ increases from $0$ to $\Delta$, $p^{\star}(x;\nu)$ is continuous and nonincreasing w.r.t. $\nu$, and is the unique solution to~\eqref{eq:wdef} for $\nu>0$.
By the KKT conditions, the optimal power maximizes the Lagrangian at the optimal dual variable $\nu^{\star}$, i.e.,
\begin{equation}\label{eq:popt}
	p[k]=p^{\star}\big(x[k];\nu^{\star}\big),\quad\forall k\in\mathcal{K},
\end{equation}
where $\nu^{\star}$ is determined by complementary slackness: $\nu^{\star}=0$ if~\eqref{eq:P1avg} is satisfied by $p^{\star}(x[k];0)$, and otherwise $\nu^{\star}$ is the root of $\frac{1}{K}\sum_{k\in\mathcal{K}}p^{\star}(x[k];\nu)=\bar P$, which can be found by bisection.

 As seen from~\eqref{eq:popt}, the optimal power depends on time only through the MA position, and is thus identical whenever the MA is at the same position.
In particular, Alice remains silent (i.e., $p[k]=0$) if and only if $x[k]$ lies in
\begin{equation}\label{eq:silence}
	\underbrace{\{x:\Delta(x)\le0\}}_{\text{Eve-dominant zone}}\cup\underbrace{\{x:0<\Delta(x)\le\nu^{\star}\ln2\}}_{\text{budget-driven zone}},
\end{equation}
which we term two-level silence: the Eve-dominant zone is independent of the average power budget $\bar P$, whereas the budget-driven zone shrinks as $\bar P$ increases.
\vspace{-5pt}
	\subsection{Trajectory Reformulation on a Discretized Rail}\label{sec:traj}
By~\eqref{eq:popt}, the objective in~\eqref{eq:P1obj} and the average power in~\eqref{eq:P1avg} depend only on the number of time slots spent at each position.
	Accordingly, we discretize the rail as in~\cite{Li2026} into $N$ grid points $x_n\triangleq(n-1)\delta$, $n\in\mathcal{N}\triangleq\{1,\dots,N\}$.
	We set the spacing to $\delta=v_{\max}\tau$, i.e., the maximum moving distance per time slot, so that the MA either stays or moves to an adjacent grid point in each time slot.\footnote{Given the $\lambda$-scale variation of the channel in~\eqref{eq:chan}, the discretization incurs an $O(v_{\max}\tau/\lambda)$ optimality loss, which vanishes as $N$ increases and is negligible since $v_{\max}\tau\ll\lambda$.}
	For simplicity, we assume that $D=(N-1)\delta$ and that the MA starts from a grid point $x_s$, i.e., $x_0=x_s$, and focus on trajectories within either $[0,x_s]$ or $[x_s,D]$.

Denote by $x_f$ the farthest grid point from $x_s$ visited by the MA.
Then, the MA passes all the $J_f+1$ grid points between $x_s$ and $x_f$ with $J_f\triangleq|f-s|$, whose indices form $\mathcal{P}_f\triangleq\{n\in\mathcal{N}:\min\{s,f\}\le n\le\max\{s,f\}\}$.
Let $m_n$ denote the visit count of $x_n$, i.e., the number of time slots that the MA spends at $x_n$, and $p_n$ the transmit power in these time slots.
For notational simplicity, the subscript $n$ denotes quantities at $x_n$, e.g., $r_n(p)\triangleq r(x_n,p)$, $w_n(\nu)\triangleq w(x_n;\nu)$, and $p_n^{\star}(\nu)\triangleq p^{\star}(x_n;\nu)$.
Thus, the objective in~\eqref{eq:P1obj} can be equivalently rewritten as
\begin{equation}\label{eq:regroup}
	\frac{1}{K}\sum_{k\in\mathcal{K}}r\big(x[k],p[k]\big)=\frac{1}{K}\sum_{n\in\mathcal{P}_f}m_nr_n(p_n),
\end{equation}
and the average power in~\eqref{eq:P1avg} equals $\frac{1}{K}\sum_{n\in\mathcal{P}_f}m_np_n$, both of which depend on the trajectory only through $f$ and $\boldsymbol m\triangleq[m_n]_{n\in\mathcal{P}_f}$.
The feasible pairs $(f,\boldsymbol m)$ are characterized as follows.

	\begin{lemma}\label{lem:traj}
	A pair $(f,\boldsymbol m)$ is attained by a trajectory satisfying~\eqref{eq:P1speed} and~\eqref{eq:P1rail} if and only if $f\in\mathcal{F}\triangleq\{f\in\mathcal{N}:J_f\le K\}$ and $\boldsymbol m\in\mathcal{M}_f$, where $\mathcal{M}_f$ is the set of integer vectors $\boldsymbol m$ with $\sum_{n\in\mathcal{P}_f}m_n=K$, $m_s\ge0$, and $m_n\ge1$ for all other $n\in\mathcal{P}_f$.
	Moreover, it suffices to consider trajectories that move from $x_s$ toward $x_f$ without turning back.
	\end{lemma}
	\begin{IEEEproof}
		Reaching $x_f$ from $x_s$ passes each $x_n$ with $n\in\mathcal{P}_f$ and $n\neq s$ at least once, which yields $m_n\ge1$ and $J_f\le K$.
		Conversely, staying at each $x_n$ for $m_n$ consecutive time slots in the order from $x_s$ to $x_f$ yields a feasible trajectory that never turns back.
	\end{IEEEproof}

Consequently, (P1) on the discretized  rail can be equivalently reformulated as
\begin{equation}\label{eq:P2}
	(\mathrm{P2}):\ \max_{f\in\mathcal{F}}\ V_f,
\end{equation}
where $V_f$ is the optimal value of
\begin{equation}\label{eq:Vf}
	\begin{aligned}
		\max_{\boldsymbol m\in\mathcal{M}_f,\,\boldsymbol p}\ &\frac{1}{K}\sum_{n\in\mathcal{P}_f}m_nr_n(p_n)\\
		\mathrm{s.t.}\ \ &\frac{1}{K}\sum_{n\in\mathcal{P}_f}m_np_n\le\bar P,\\
		&0\le p_n\le P_{\max},\ \forall n\in\mathcal{P}_f,
	\end{aligned}
\end{equation}
with $\boldsymbol p\triangleq[p_n]_{n\in\mathcal{P}_f}$.
Although the movement constraints become linear in $\boldsymbol m$ for a given $f$, \eqref{eq:Vf} is still a nonconvex mixed-integer problem, since $m_n$ multiplies $r_n(p_n)$ and $p_n$ in the objective and the power constraint, respectively. This difficulty is tackled in Section~\ref{sec:sol} via convex relaxation and Lagrangian duality.

\vspace{-3pt}
	\section{Dual-Based Branch-and-Bound Algorithm}\label{sec:sol}
	\subsection{Convex Relaxation and Lagrangian Dual}\label{sec:dual}
For a given $f$, we relax $\mathcal{M}_f$ into its continuous counterpart $\bar{\mathcal{M}}_f$ and replace each power $p_n$ with the energy $\varepsilon_n\triangleq m_np_n$, which transforms~\eqref{eq:Vf} into
\begin{equation}\label{eq:Rf}
	\begin{aligned}
		(\mathrm{P3}):\ \bar V_f\triangleq\max_{(\boldsymbol m,\boldsymbol\varepsilon)\in\mathcal{Z}_f}\ &\frac{1}{K}\sum_{n\in\mathcal{P}_f}\varphi_n(m_n,\varepsilon_n)\\[-1pt]
		\mathrm{s.t.}\ \ &\frac{1}{K}\sum_{n\in\mathcal{P}_f}\varepsilon_n\le\bar P,
	\end{aligned}
\end{equation}
where $\mathcal{Z}_f\triangleq\{(\boldsymbol m,\boldsymbol\varepsilon):\boldsymbol m\in\bar{\mathcal{M}}_f,\,0\le\varepsilon_n\le m_nP_{\max},\,\forall n\in\mathcal{P}_f\}$, and $\varphi_n(m,\varepsilon)\triangleq m\,r_n(\varepsilon/m)$ is the perspective function of $r_n$, with $\varphi_n(0,0)\triangleq0$.
Since $r_n$ is concave, $\varphi_n$ is jointly concave in $(m_n,\varepsilon_n)$~\cite{Boyd2004}. Hence, (P3) is a convex problem, and $\bar V_f\ge V_f$ holds due to the relaxation.

We solve (P3) via Lagrangian duality, since its dual function has a closed form  that also provides upper bounds on $V_f$ in (P2).
As in Section~\ref{sec:power}, we introduce a dual variable $\nu\ge0$ for the power constraint of (P3).
For a given $\boldsymbol m$, maximizing the Lagrangian over each $\varepsilon_n=m_np_n$ is equivalent to maximizing $m_n[r_n(p_n)-\nu p_n]$ over $p_n\in[0,P_{\max}]$, which yields $m_nw_n(\nu)$ at $p_n=p_n^{\star}(\nu)$ by~\eqref{eq:wdef}.
Since the resulting objective is linear in $\boldsymbol m$, it is maximized by assigning one time slot to each $x_n$ with $n\in\mathcal{P}_f$, $n\neq s$, and the remaining $K-J_f$ time slots to the grid point with the largest net secrecy gain, given by $W_f(\nu)\triangleq\max_{n\in\mathcal{P}_f}w_n(\nu)$.
Accordingly, the dual function of (P3) is given by
\begin{equation}\label{eq:gf}
	g_f(\nu)=\frac{1}{K}\Big[\sum_{n\in\mathcal{P}_f,\,n\neq s}w_n(\nu)+(K-J_f)\,W_f(\nu)\Big]+\nu\bar P.
\end{equation}
By weak duality, $g_f(\nu)\ge\bar V_f\ge V_f$ holds for all $\nu\ge0$.
Moreover, since (P3) is convex and strictly feasible, strong duality holds by Slater's condition~\cite{Boyd2004}, i.e., $\bar V_f=\min_{\nu\ge0}g_f(\nu)$.

\vspace{-8pt}
\subsection{Primal Solution Recovery}\label{sec:dwell}
To recover the primal solution from~\eqref{eq:gf}, note that $W_f(\nu)$ may be attained by multiple grid points, which form the set of candidate dwell points $\mathcal{A}_f(\nu)\triangleq\arg\max_{n\in\mathcal{P}_f}w_n(\nu)$.
If all the remaining $K-J_f$ time slots are assigned to $x_n$ with $n\in\mathcal{A}_f(\nu)$, the resulting average power is $\bar p_{f,n}(\nu)\triangleq\frac{1}{K}\Big[\sum_{i\in\mathcal{P}_f,\,i\neq s}p_i^{\star}(\nu)+(K-J_f)\,p_n^{\star}(\nu)\Big]$, based on which an optimal solution to (P3) is given as follows.

	\begin{proposition}\label{prop:dwell}
		For the optimal dual variable $\nu_f^{\star}\in\arg\min_{\nu\ge0}g_f(\nu)$, let $n^{+},n^{-}\in\mathcal{A}_f(\nu_f^{\star})$ maximize and minimize $p_n^{\star}(\nu_f^{\star})$, respectively.
		Then, (P3) has an optimal solution with $\varepsilon_n=m_np_n^{\star}(\nu_f^{\star})$ and
\begin{equation}\label{eq:mstar}
	m_n=\ind_{\{n\neq s\}}+(K-J_f)\big[\rho\,\ind_{\{n=n^{+}\}}+(1-\rho)\,\ind_{\{n=n^{-}\}}\big],
\end{equation}
where $\rho=0$ if $\nu_f^{\star}=0$, and otherwise $\rho\in[0,1]$ satisfies
\begin{equation}\label{eq:rho}\rho\,\bar p_{f,n^{+}}(\nu_f^{\star})+(1-\rho)\,\bar p_{f,n^{-}}(\nu_f^{\star})=\bar P.\end{equation}
	\end{proposition}
	\begin{IEEEproof}
		Please see the Appendix.
	\end{IEEEproof}

Proposition~\ref{prop:dwell} implies that the MA dwells at no more than two grid points and otherwise moves toward $x_f$ at the maximum speed. 
Since $\rho(K-J_f)$ is generally fractional, we round it down by moving its fractional part (denoted by $\eta\in[0,1)$) from $x_{n^{+}}$ to $x_{n^{-}}$, which yields a feasible $\hat{\boldsymbol m}\in\mathcal{M}_f$. 
As $p_{n^{-}}^{\star}(\nu_f^{\star})\le p_{n^{+}}^{\star}(\nu_f^{\star})$, this move keeps the power budget satisfied and changes the objective by $\frac{\eta}{K}\big[r_{n^{-}}(p_{n^{-}}^{\star}(\nu_f^{\star}))-r_{n^{+}}(p_{n^{+}}^{\star}(\nu_f^{\star}))\big]\ge-\bar R/K$, where $\bar R\triangleq\max_{n\in\mathcal{N}}r_n(P_{\max})$. 
Then, we optimize the transmit powers for the resulting trajectory via~\eqref{eq:popt}, which does not decrease the objective. Therefore, the achieved value $\hat V_f$ satisfies
\begin{equation}\label{eq:round}
	\bar V_f-\bar R/K\le\hat V_f\le V_f\le\bar V_f,
\end{equation}
which implies asymptotic optimality as $K\to\infty$.

\begin{algorithm}[!t]
	\caption{Dual-Based Branch-and-Bound (BnB)}\label{alg:bnb}
	\small\begin{algorithmic}[1]\raggedright
		\Require $\{\alpha_n,\beta_n\}_{n\in\mathcal{N}}$, $s$, $K$, $\bar P$, $P_{\max}$, coarse grid $\Omega$ of $\nu$
		\State Initialize $\bar G_f\gets\min_{\nu\in\Omega}g_f(\nu)$ via~\eqref{eq:gf}, $\forall f\in\mathcal{F}$
		\State Sort $\mathcal{F}$ in descending order of $\bar G_f$; set $\mathrm{LB}\gets-\infty$
		\For{each $f\in\mathcal{F}$}
	   \State \textbf{if} $\bar G_f\le\mathrm{LB}$ \textbf{then} skip $f$  \Comment{Prune}
		\State Find $\nu_f^{\star}\in\arg\min_{\nu\ge0}g_f(\nu)$ by bisection and $n^{\pm}$ via Prop.~\ref{prop:dwell}
		\State Update $\bar G_i\gets\min\{\bar G_i,g_i(\nu_f^{\star})\}$, $\forall i\in\mathcal{F}$
		\State Round~\eqref{eq:mstar} into $\hat{\boldsymbol m}$ and obtain the optimal powers via~\eqref{eq:popt}
		\State Update $\mathrm{LB}\gets\max\{\mathrm{LB},\hat V_f\}$ and the best solution
		\EndFor
		\Ensure Best solution with value $\mathrm{LB}$; $\mathrm{UB}\gets\max_{f\in\mathcal{F}}\bar G_f$
	\end{algorithmic}
\end{algorithm}
\vspace{-5pt}
	\subsection{Branch-and-Bound Search and Complexity}\label{sec:alg}
Based on the above results, it remains to find the optimal farthest grid point in (P2).
Instead of exhaustively solving (P3) for every $f\in\mathcal{F}$, we develop a low-complexity dual-based BnB algorithm that prunes unpromising candidates via the dual upper bounds in~\eqref{eq:gf}, as summarized in Algorithm~\ref{alg:bnb}.
Specifically, the upper bound $\bar G_f$ of each $f\in\mathcal{F}$ is initialized as the minimum of $g_f(\nu)$ over a coarse grid $\Omega$ of $\nu$.
For each $\nu\in\Omega$, $g_f(\nu)$ for all $f\in\mathcal{F}$ is obtained with only $O(N)$ operations, since the sum and the maximum in~\eqref{eq:gf} can be computed as prefix sums and prefix maxima.
The candidates are then examined in descending order of $\bar G_f$, and $f$ is pruned if $\bar G_f\le\mathrm{LB}$, where $\mathrm{LB}$ is the best objective value found so far.
 Otherwise, we find $\nu_f^{\star}$ by bisection, tighten all bounds via $g_i(\nu_f^{\star})$, and construct a feasible solution by rounding~\eqref{eq:mstar}.
Let $V^{\star}$ denote the optimal value of (P2), and $\mathrm{UB}$ the largest final bound $\max_{f\in\mathcal{F}}\bar G_f$.
By~\eqref{eq:round} and the pruning rule, the output solution satisfies $V^{\star}-\bar R/K\le\mathrm{LB}\le V^{\star}\le\mathrm{UB}$, which certifies its optimality gap.

The overall complexity of Algorithm~\ref{alg:bnb} is $O\big((|\Omega|+N_{\mathrm{r}}({I_{\mathrm{d}}+I_{\mathrm{p}}}))N\big)$, where $N_{\mathrm{r}}$ is the number of unpruned candidates, and $I_{\mathrm{d}}$ and $I_{\mathrm{p}}$ are the numbers of iterations of the bisections in Steps~5 and~7, respectively.
For a given $N_{\mathrm{r}}$, this complexity is linear in $N$, whereas applying the DP in~\cite{Li2026} to the same grid requires $O(KN)$ operations.

\begin{remark}\label{rem:dp}
Extending the DP in~\cite{Li2026} to (P1) by dualizing~\eqref{eq:P1avg} requires executing the DP for each value of the dual variable, and the resulting solution is not guaranteed to be optimal due to the potential duality gap of (P1)~\cite{Yu2006}.
\end{remark}

\vspace{-5pt}
	\section{Simulation Results}\label{sec:sim}
\begin{figure}[!t]
		\centering
		\includegraphics[width=1\columnwidth]{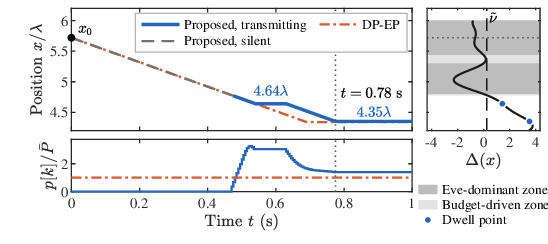}
		\caption{Proposed design versus DP-EP in one realization. Left: position and normalized power versus time. Right: $\Delta(x)$ on the same position axis.}

		\label{fig:mech}
	\end{figure}

Unless otherwise specified, we set the carrier wavelength $\lambda=0.06$~m, the  rail length $D=6\lambda$, and the maximum speed $v_{\max}=0.12$~m/s.
The block duration is $T=1$~s and the time slot length is $\tau=5$~ms (i.e., $K=200$), yielding $\delta=v_{\max}\tau=\lambda/100$ and $N=601$.
    The initial position $x_0$ is randomly selected from the grid points for each channel realization.
    Bob's and Eve's channels each consist of $L_{\mathrm{b}}=L_{\mathrm{e}}=6$ paths, whose AoDs are uniformly distributed over $[0,\pi]$ and whose gains follow $a_{u,l}\sim\mathcal{CN}(0,\mathrm{PL}(d_u)/L_u)$, where $\mathrm{PL}(d)\triangleq(\lambda/(4\pi))^2d^{-2.8}$ and the Alice--Bob and Alice--Eve distances are $d_{\mathrm{b}}=d_{\mathrm{e}}=50$~m.
	The noise powers are $\sigma_{\mathrm{b}}^2=\sigma_{\mathrm{e}}^2=-90$~dBm, and the average and peak transmit powers are $\bar P=10$~dBm and $P_{\max}=\bar P+6$~dB, respectively.
	Except for Fig.~\ref{fig:mech}, all results are averaged over 500 channel realizations.

\begin{table}[!t]
	\centering
	\caption{Average Secrecy Rate, Optimality Gap, and Runtime}\label{tab:cplx}
	\vspace{-4pt}
	\footnotesize
	\setlength{\tabcolsep}{2.6pt}
	\begin{tabular}{ccc|c|ccc}
		\hline
		\multirow{2}{*}{\(\tau\) (ms)} & \multirow{2}{*}{\(K/N\)} & \multirow{2}{*}{\(\bar C\) (bps/Hz)} & \multirow{2}{*}{\(\frac{\mathrm{UB}-\mathrm{LB}}{\mathrm{UB}}\)} & \multicolumn{3}{c}{Runtime (ms)} \\
		& & & & Proposed & Dual-DP & DP-EP \\
		\hline
		5     & 600/601   & 2.3110 & \(2.7\times10^{-7}\) & 1.8 & 146 & 3.3 \\
		2.5   & 1200/1201 & 2.3111 & \(3.8\times10^{-8}\) & 2.8 & 349 & 8.5 \\
		1.25  & 2400/2401 & 2.3112 & \(3.7\times10^{-8}\) & 4.8 & 969 & 25.4 \\
		\hline
	\end{tabular}
\end{table}

   For comparison, the following benchmark schemes are considered, all of which start from $x_0$ and satisfy the constraints in (P1):
	1)~DP-EP (equal power): the DP in~\cite{Li2026} with the per-slot secrecy reward $r(x,\bar P)$ and a constant power $\bar P$;
	2)~DP-Rate: the original design of~\cite{Li2026}, i.e., the DP with the rate reward $\log_2(1+\alpha(x)\bar P)$ and a constant power $\bar P$;
	3)~RTT-DA: a delay-aware (DA) relocate-then-transmit (RTT) scheme, which stays silent while moving to a dwell position optimized under the movement delay, and then spreads the energy budget evenly over the remaining time slots (capped at $P_{\max}$);
    4)~RTT-DU: a delay-unaware (DU) version of RTT-DA;
	5)~FPA: a fixed-position antenna at $x_0$ with constant power $\bar P$.

	\begin{figure*}[!t]
	\centering
	\includegraphics[width=\textwidth]{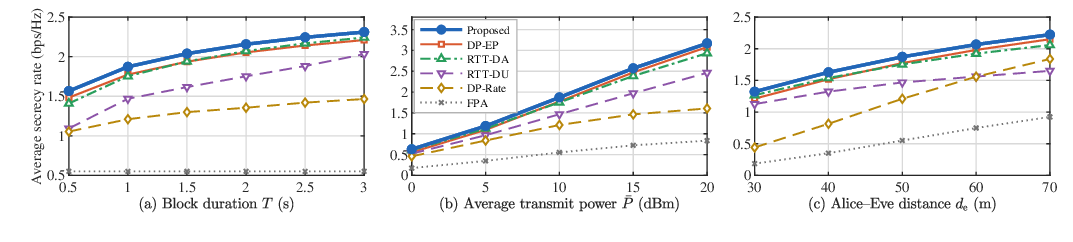}
	\caption{Average secrecy rate versus (a) the block duration $T$, (b) the average transmit power $\bar P$, and (c) the Alice--Eve distance $d_{\mathrm e}$.}
	\label{fig:res}
	\vspace{-12pt}
\end{figure*}

    Table~\ref{tab:cplx} lists the average secrecy rate $\bar C$, the relative optimality gap, and the runtimes for different $\tau$ with $T=3$~s, where Dual-DP denotes the extended DP in Remark~\ref{rem:dp}.
	As $\tau$ decreases, $\bar C$ remains virtually unchanged and the optimality gaps are below $10^{-6}$, which demonstrates that the discretization loss is already negligible at $\tau=5$~ms and the proposed algorithm is essentially optimal.
    Notably, the proposed algorithm solves the joint problem even faster than DP-EP, and this advantage widens as $K$ increases, which corroborates the analysis in Section~\ref{sec:alg}.
    Meanwhile, Dual-DP is about two orders of magnitude slower than the proposed algorithm and is thus excluded hereafter.

     Fig.~\ref{fig:mech} illustrates a realization with $x_0=5.72\lambda$.
	 DP-EP stops at $4.34\lambda$ and transmits at constant power. In contrast, the proposed design moves at $v_{\max}$ while keeping silent in the Eve-dominant and budget-driven zones, and dwells first at $4.64\lambda$ and then at $4.35\lambda$.
    Consistent with Proposition~\ref{prop:dwell}, the two dwell points yield identical net secrecy gains, and the remaining time slots are split between them to meet the average power budget.

     Fig.~\ref{fig:res}(a) depicts the average secrecy rate versus the block duration $T$.
FPA performs the worst, since it can exploit neither the spatial channel variation nor power adaptation.
    RTT-DU and DP-Rate are inferior to RTT-DA and DP-EP, as they ignore the movement delay and Eve, respectively.
    For large $T$, RTT-DA even outperforms DP-EP, since it has enough time to spend its entire energy budget at a favorable position, whereas DP-EP must spend $\bar P$ even in the Eve-dominant zone while moving.
	Nevertheless, the proposed design achieves a higher average secrecy rate than both RTT-DA and DP-EP across the considered range of $T$.

	Fig.~\ref{fig:res}(b) shows the average secrecy rate versus the average transmit power $\bar P$.
At low $\bar P$, RTT-DA closely approaches the proposed design, since the budget-driven zone expands as $\bar P$ decreases, and thus the optimal strategy concentrates energy at the most favorable dwell points.
	At high $\bar P$, since the secrecy rate at each position saturates at $\log_2(\alpha(x)/\beta(x))$, spreading the power over more time slots becomes more effective, which enables DP-EP to outperform RTT-DA.
    Overall, the proposed design  outperforms both RTT-DA and DP-EP over the considered power range.

	Fig.~\ref{fig:res}(c) illustrates the average secrecy rate versus the Alice--Eve distance $d_{\mathrm e}$.
For $d_{\mathrm e}<d_{\mathrm b}$, RTT-DA outperforms DP-EP, since the Eve-dominant zones expand and penalize the continuous transmission of DP-EP. Conversely, DP-EP outperforms RTT-DA for $d_{\mathrm e}>d_{\mathrm b}$, as these zones shrink.
	Nevertheless, the proposed design consistently outperforms all benchmarks.


	\section{Conclusion}
  This paper investigated joint trajectory and power optimization for MA-aided secure transmission, where the MA can transmit during its movement.
 By exploiting the position-dependent optimal power, we equivalently reformulated the problem on a discretized  linear rail in terms of the farthest grid point and the number of time slots spent at each grid point. Based on the convex relaxation of this problem, we developed a low-complexity dual-based BnB algorithm that yields an asymptotically optimal solution with a certified optimality gap.
  We also showed that the MA dwells at no more than two grid points and otherwise moves at the maximum speed, remaining silent in both the Eve-dominant and budget-driven zones.
  Simulation results showed that the proposed design outperforms the benchmark schemes and requires less computation time than the DP-based methods.

	\appendix
	\begin{IEEEproof}[Proof of Proposition~\ref{prop:dwell}]
	As the pointwise supremum of affine functions of $\nu$, $w_n(\nu)$ is convex.
    Since the maximizer in~\eqref{eq:pstar} is unique for $\nu>0$, Danskin's theorem gives $w_n'(\nu)=-p_n^{\star}(\nu)$, and hence $\partial W_f(\nu)=\mathrm{conv}\{-p_n^{\star}(\nu)\}_{n\in\mathcal{A}_f(\nu)}$.
	 Substituting this into~\eqref{eq:gf} yields $\partial g_f(\nu)=\big\{\bar P-q:\ q\in\mathrm{conv}\{\bar p_{f,n}(\nu)\}_{n\in\mathcal{A}_f(\nu)}\big\}$, which reduces to the interval $[\bar P-\bar p_{f,n^{+}}(\nu_f^{\star}),\,\bar P-\bar p_{f,n^{-}}(\nu_f^{\star})]$ at $\nu=\nu_f^{\star}$.

	 If $\nu_f^{\star}>0$, the optimality condition $0\in\partial g_f(\nu_f^{\star})$ implies $\bar P\in[\bar p_{f,n^{-}}(\nu_f^{\star}),\,\bar p_{f,n^{+}}(\nu_f^{\star})]$, which guarantees the existence of $\rho\in[0,1]$ satisfying~\eqref{eq:rho}.
	Then, the solution defined by~\eqref{eq:mstar} and~\eqref{eq:rho} meets the power constraint with equality, and is thus primal feasible and satisfies complementary slackness.
	If $\nu_f^{\star}=0$, the optimality condition requires $g_f'(0^{+})=\bar P-\bar p_{f,n^{-}}(0)\ge0$, and thus the solution with $\rho=0$ is primal feasible and trivially satisfies complementary slackness.
    In both cases, the solution also maximizes the Lagrangian at $\nu_f^{\star}$, and is thus optimal for (P3) by strong duality.
	\end{IEEEproof}


\end{document}